\documentclass[a4paper,11pt]{article}
\usepackage{pos}

\title{Numerical study of the dimensionally reduced 3D Ising model}
\author*[a]{Tolga Kiel}
\author[a,b]{Stephan D\"urr}

\affiliation[a]{Physics Department, University of Wuppertal, 42119 Wuppertal, Germany}
\affiliation[b]{IAS/JSC, Forschungszentrum J\"ulich, 52425 J\"ulich, Germany}

\emailAdd{tolga.kiel\,(AT)\,uni-wuppertal$\mbox{.}$de}
\emailAdd{duerr\,(AT)\,uni-wuppertal$\mbox{.}$de}
\abstract{We study the 3D Ising model in the infinite volume limit
$N_{x,y,z}\to\infty$ by means of numerical simulations. We determine $T_c$
as well as the critical exponents $\beta,\gamma$ and $\nu$, based
on finite-size scaling and histogram reweighting techniques. In addition,
we study a ``dimensionally reduced'' scenario where $N_z$ is kept fixed
(e.g.\ at 2, 4, 8), while the limit $N_{x,y}\to\infty$ is taken. For each
fixed $N_z$ we determine $T_c$ as well as $\beta,\gamma,\nu$. For
$T_c$ we find a smooth transition curve which connects the well known
critical temperatures of the 2D and the 3D Ising model. Regarding
$\beta,\gamma,\nu$ our data suggest that the ``dimensionally
reduced'' Ising model is in the same universality class as the 2D Ising
model, regardless of $N_z$.}

\FullConference{%
  41st International Symposium on Lattice Field Theory\\
  July 28, 2024 to August 3, 2024\\
  University of Liverpool, United Kingdom 
}

\begin{document}
\maketitle

\section{Introduction}
The Ising model is analytically solvable in 2D \cite{Isi25}, and it has been
investigated on various occasions in 3D. We would like to know whether the
two models connect smoothly to each other if one studies a dimensionally
reduced version of the latter, i.e.\ a model on a $N_x \times N_y \times
N_z$ lattice where only the extensions $N_x=N_y$ are taken large (jointly
dubbed $L$ below), while $N_z$ is kept fixed.\\

To this end we write a 3D Ising model code with periodic boundary conditions, so that we may study the 3D Ising model in the infinite volume limit, $N_{x,y,z} \to \infty$, as well as the dimensionally reduced Ising model with fixed $N_z=1,2,4,8$, in the limit $N_{x,y} \to \infty$, using Monte Carlo simulations. Our goal is to determine the critical temperature $T_c$ and the critical exponents $\beta$, $\gamma$, and $\nu$ through finite-size scaling and histogram reweighting techniques. Simulations are performed for the ferromagnetic Ising model, governed by the Hamiltonian and partition function
\begin{align}\label{Hamiltonian}
    \mathcal{H} = -J \sum_{\langle ij\rangle} \sigma_i \sigma_j,&& \mathcal{Z}=\sum_{\mathrm{configs}\{\sigma\}} e^{-\beta\mathcal{H}}
\end{align}
with $J>0$. We define the dimensionless coupling $\tilde{J}:=\beta J$ with $\beta=1/(k_\mathrm{B}T)$. The spin variables $\sigma_i$ take values $\pm1$ and the sum is over nearest-neighbor pairs $\langle ij\rangle$, where $i$, $j$ label sites in 2D or 3D.

\section{Numerical methods}
\subsection{Monte Carlo sampling method}
We simulate $L^3$ and $L\times L \times N_z$ Ising lattices using the Metropolis algorithm in combination with the Wolff cluster algorithm. This way we ensure that the code is reasonably efficient, regardless whether the dialed parameter $\tilde{J}$ is close to $\tilde{J}_c$ or far away from the latter. Random numbers are generated using George Marsaglia's KISS random number generator. For the 3D Ising model, we obtain data for box sizes $L$ with $24\leq L\leq 256$. In case of the dimensionally reduced Ising model with $N_z=1,2,4,8$, we use various ranges of $L$, for instance $32\leq L \leq 2048$ for $N_z=8$. We perform $\mathcal{O}(10^6 $-$ 10^8)$ measurements with 10 updates between adjacent measurements, where an update is defined as a Metropolis sweep over the whole lattice, followed by a Wolff cluster update. $\mathcal{O}(10^5)$ measurements are discarded for thermalization before data acquisition begins.

\subsection{Observables}
We measure the following observables for a system with $N = N_x N_y N_z$ sites
\begin{align}
    m = \frac{1}{N} \sum_{ i \in \Lambda} \sigma_i && \chi = JN(\langle |m|^2 \rangle - \langle |m| \rangle^2) && U_4 = 1 - \frac{\langle |m| \rangle^4}{3\langle |m|^2 \rangle^2}
\end{align}
where $\langle . \rangle$ denotes the ensemble average, $m$ is called the magnetization, $\chi$ the magnetic susceptibility and $U_4$ the fourth-order Binder cumulant of the magnetization \cite{Binder:1981sa}. Both in $\chi$ and $U_4$ we use the finite-volume version ($|m|$ instead of $m$).
\subsection{Finite-size scaling}
We use the finite size scaling theory, first developed by Fisher \cite{Fisher72,Priv91,Binder}, to determine the critical exponents. For large values of $L$, the following scaling relations are expected to hold
\begin{align}\label{eq:scaling-relations}
    \left.\langle |m| \rangle\right\vert_{\Tilde{J} = \Tilde{J_c}} \propto L^{-\beta/\nu}&&\max_{\tilde{J}} \chi \propto L^{\gamma/\nu}&&
    \left.\frac{\partial U_4}{\partial \Tilde{J}}\right\vert_{\Tilde{J} = \Tilde{J_c}} \propto L^{1/\nu}.
\end{align}
To determine the critical coupling $\tilde{J}_c$ we use Binder's fourth-order cumulant crossing technique. As the lattice size $L\to\infty$, the Binder cumulant $U_4\to0$ for $\Tilde{J}<\Tilde{J}_c$ and $U_4\to 2/3$ for $\Tilde{J}>\Tilde{J}_c$. One can plot $U_4$ as a function of $\tilde{J}$ for different lattice sizes. For large enough values of $L$, the locations of the intersections indicate $\tilde{J}_c$.

\subsection{Histogram reweighting}
The use of histograms allows to obtain additional information from Monte Carlo simulations by transforming samples from a known probability distribution into samples from a different distribution within the same state space \cite{FerSwe88,FerSwe89}.
A Monte Carlo simulation is first run at the inverse temperature $\tilde{J}^\prime$. The expectation value of an observable $\mathcal{O}$ at another coupling $\tilde{J}$ in the vicinity of $\tilde{J}^\prime$ can be determined via
\begin{equation}\label{Reweighting formula}
    \langle \mathcal{O} \rangle_{\Tilde{J}} = \frac{\left \langle \mathcal{O} e^{-(\Tilde{J} - \Tilde{J}^\prime)E} \right \rangle_{\Tilde{J}^\prime}}{\left \langle e^{-(\Tilde{J} - \Tilde{J}^\prime)E} \right \rangle_{\Tilde{J}^\prime}}.
\end{equation}
As $\tilde{J}$ can be varied continuously, the histogram method is able to precisely locate the peak in $\chi$ and the intersections of $U_4$.

\subsection{Estimation of peak parameters}
Because of the exponential increase in statistical errors when reweighting to a coupling $\tilde{J}$ significantly different from the simulated coupling $\tilde{J}^\prime$, histogram reweighting is only feasible in close proximity to $\tilde{J}^\prime$. By fitting data from multiple simulations performed around the estimated peak of the magnetic susceptibility $\chi$ or in the vicinity of the critical coupling $\tilde{J}_c$, we get preliminary estimates of the relevant couplings for further simulations.

\begin{figure}[hbt!]
    \centering
    \includegraphics[width=0.496\linewidth]{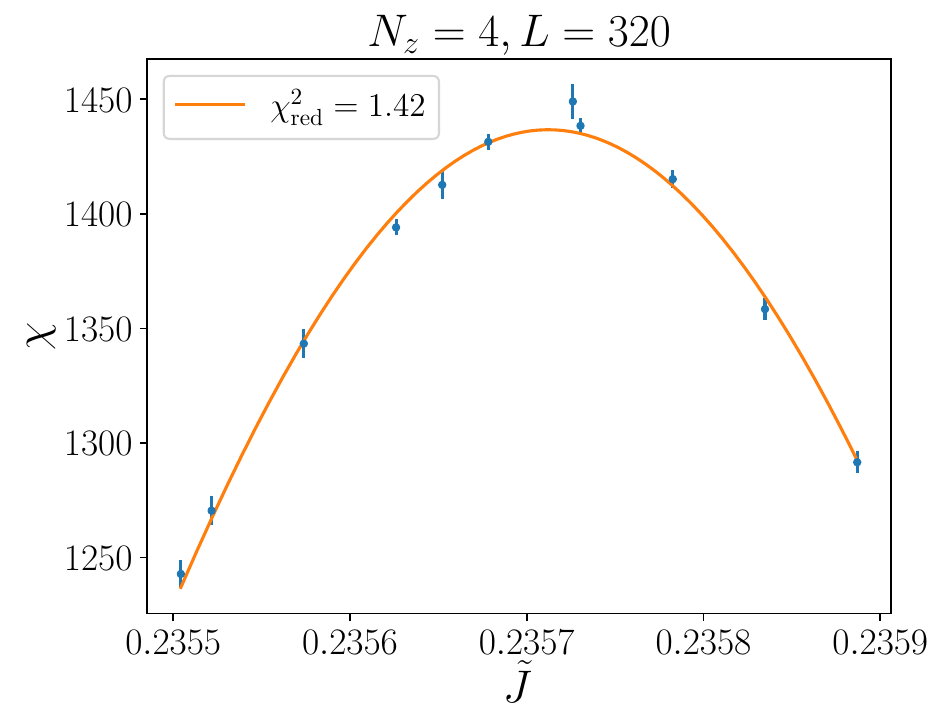}
    \includegraphics[width=0.485\linewidth]{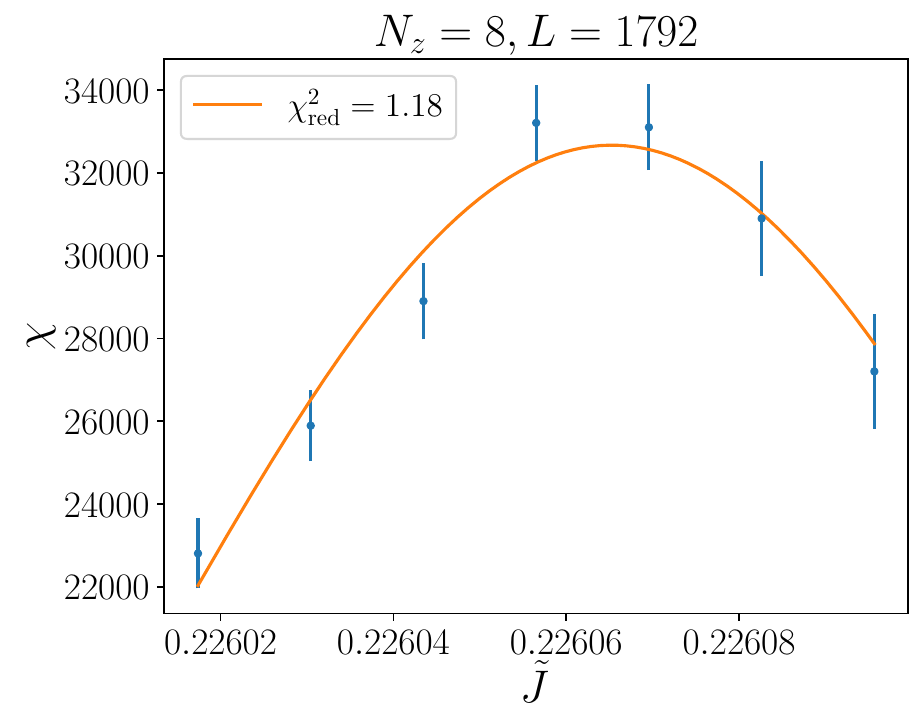}
    \vspace*{-3mm}
    \caption{Gaussian fit to the peak region of the magnetic susceptibility for $N_z=4$, $L=320$ (left) and $N_z=8$, $L=1792$ (right). The locations of the peaks are used as preliminary estimates for further simulations.}\label{fig:peaks}
    \label{fig:2DCrossing}
\end{figure}

\begin{figure}[hbt!]
    \centering
    \includegraphics[width=\linewidth]{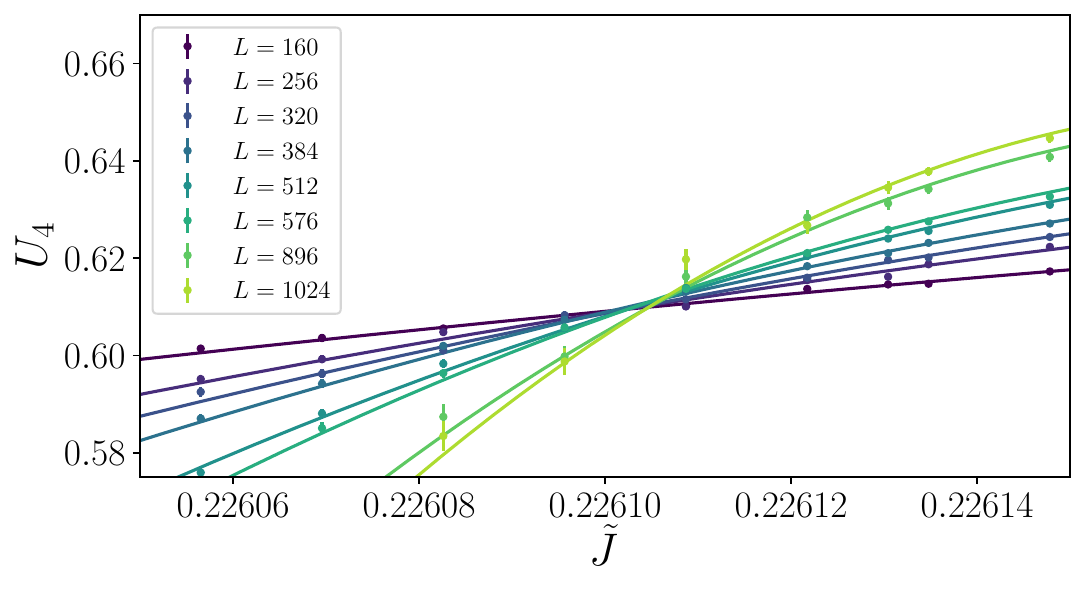}
    \vspace*{-8mm}
    \caption{Binder cumulant $U_4$ of the magnetization versus $\tilde{J}$ for $L \times L \times 8$ Ising lattices. The curves show quadratic fits to the data, used to determine a preliminary estimate of the critical coupling at $N_z=8$.}
    \label{fig:Binder_cumulant_preliminary}
\end{figure}

Fig.~\ref{fig:peaks} shows Gaussian fits to the peak regions of $\chi$ for $N_z=4$, $L=320$ in the left panel and $N_z=8$, $L=1792$ in the right panel. Fig.~\ref{fig:Binder_cumulant_preliminary} presents quadratic fits to $U_4$ for a selection of $L\times L\times 8$ lattices. By averaging the intersection points of $U_4$ for various $L$, a preliminary estimate of $\tilde{J}_c$ is obtained for fixed $N_z$ (here $N_z=8$).

The couplings determined from the peaks in $\chi$ (as identified by the fits) are used to perform additional simulations, which provide the final results for $\gamma/\nu$. Similarly, the preliminary estimates of $\Tilde{J}_c$ serve as the couplings for the simulations used to obtain the results for $\tilde{J}_c$, $\beta/\nu$ and $\nu$.

\newpage
\subsection{Error analysis}
We perform a delete-$d$-jackknife analysis to estimate the statistical errors of all quantities, where $d$ is chosen such that the data is divided into $10000$ jackknife-blocks. Because of ensemble sizes of $\mathcal{O}(10^6 $-$ 10^8)$ and integrated autocorrelation times $1\leq\tau_{\mathrm{int}}\leq15$ (in original units), it is ensured that $d\gg2\tau_{\mathrm{int}}+1$.
In addition to statistical errors, one also has to deal with systematic errors which stem from the fact that scaling relations are only valid for asymptotically large $L$. Our strategy is to exclude the smallest systems from the analysis one by one, until the estimator of the desired quantity, which includes only data with $L_{\mathrm{min}}\leq L$, does not change significantly any more.

\section{Results}
\subsection{Critical coupling}
\begin{figure}[hbt!]
    \centering
    \includegraphics[width=0.485\linewidth]{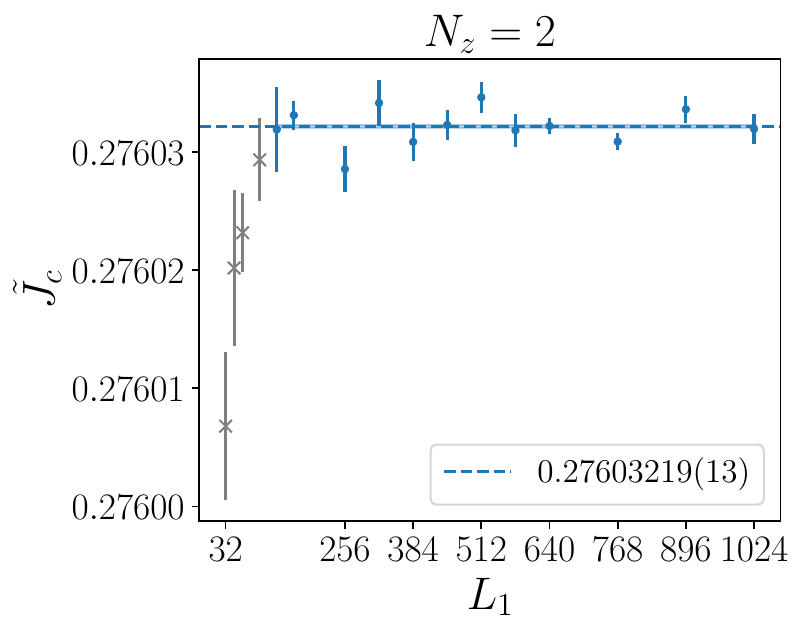}
    \includegraphics[width=0.485\linewidth]{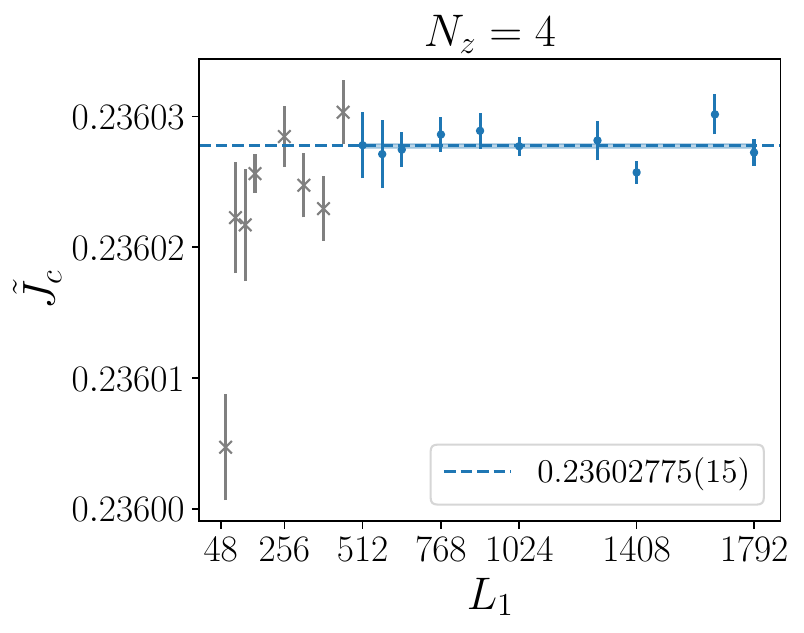}
    \vspace*{-3mm}
    \caption{Locations of the Binder cumulant crossing of Fig.~\ref{fig:Binder_cumulant_preliminary} for pairs of increasing lattice sizes $L_1<L_2$ of two $L_i \times L_i \times N_z$
    geometries ($i=1$ or $i=2$) at $N_z=2$ (left) and $N_z=4$ (right).}
    \label{fig:binder_crossing}
\end{figure}

\begin{figure}[hbt!]
    \centering
    \includegraphics[width=0.9\linewidth]{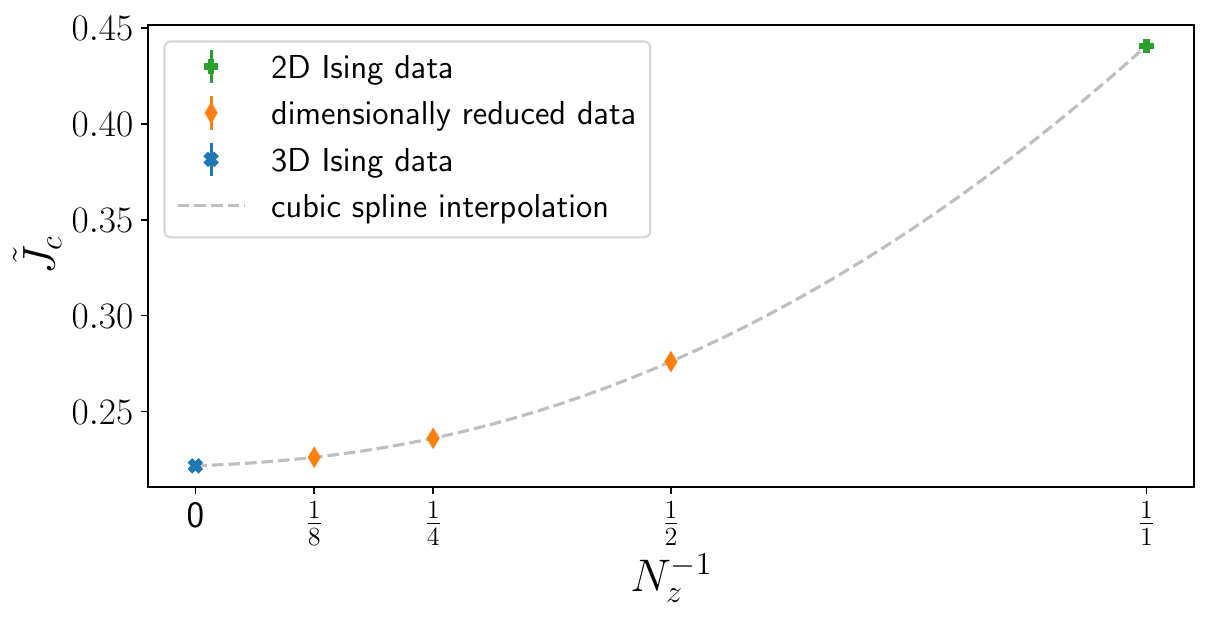}
    \vspace*{-3mm}
    \caption{Critical couplings $\Tilde{J}_c$ as a function of $N_z^{-1}$. A cubic spline interpolation is shown to guide the eye. Error bars are smaller than the symbol size.}
    \label{fig:Jc_Nz}
\end{figure}

Fig.~\ref{fig:binder_crossing} shows the locations of the Binder cumulant crossings for pairs of increasing lattice sizes $L_1<L_2$ of two $L_i \times L_i \times N_z$ geometries ($i=1$ or $i=2$) at $N_z=2$ and $N_z=4$. One can clearly see the systematic deviation for small $L_1$. The location of the intersection seems to reach a plateau at $L_1=128$ in the left panel and $L_1=512$ in the right panel. To obtain an estimator of the critical coupling $\tilde{J}_c$ for a fixed $N_z$, we calculate the weighted average of all crossings where $L_1$ reaches the respective plateau. Fig.~\ref{fig:Jc_Nz} shows $\tilde{J}_c$ as a function of $N_z^{-1}$, together with a cubic spline interpolation to guide the eye. One can clearly see a smooth transition between the couplings of the 2D Ising model and the 3D Ising model. The latter has been investigated in Refs.~\cite{Caselle:1995wn,Suarez2015,Phu2009,Ferrenberg:2018zst}.

\subsection{Critical exponents $\beta$, $\gamma$, $\nu$ and effective dimension $d_{\mathrm{eff}}$}
\begin{figure}[hbt!]
    \centering
    \includegraphics[width=0.45\linewidth]{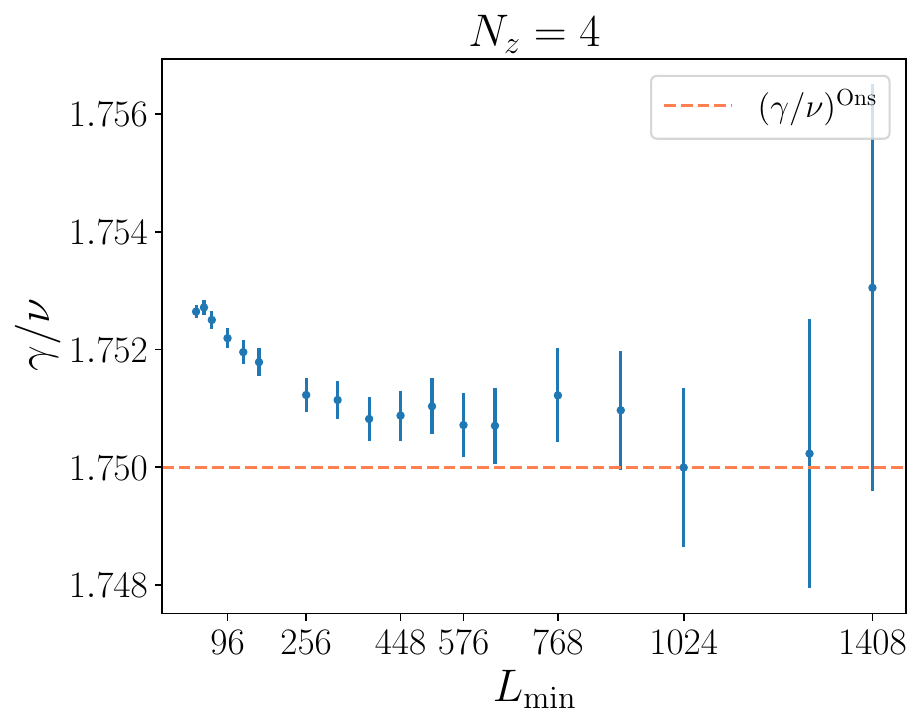}
    \includegraphics[width=0.45\linewidth]{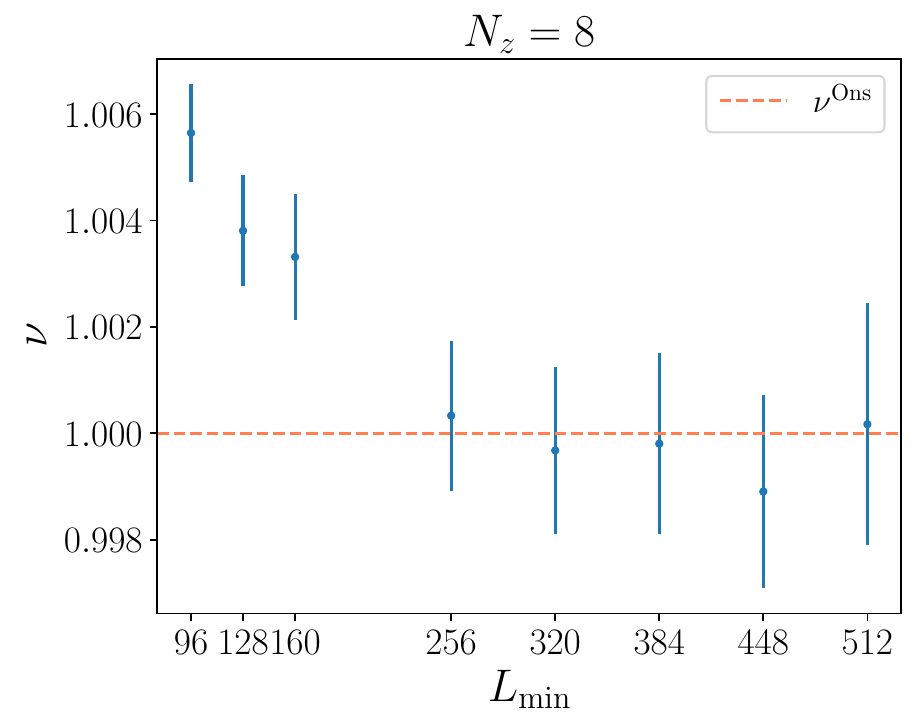}
    \vspace*{-3mm}
    \caption{$L_\mathrm{min}$ dependence of the estimate of the critical exponent
    $\gamma/\nu$ for $N_z=4$ (left) and of $\nu$ for $N_z=8$ (right). The estimators seem to reach a plateau (with our error bars) at
    $L_\mathrm{min}=448$ and $L_\mathrm{min}=256$, respectively. The
    analytic values of the 2D Ising model are shown as dashed lines for comparison.}\label{fig:exp_scan}
\end{figure}

\begin{figure}[hbt!]
     \centering
     \includegraphics[width=0.46\linewidth]{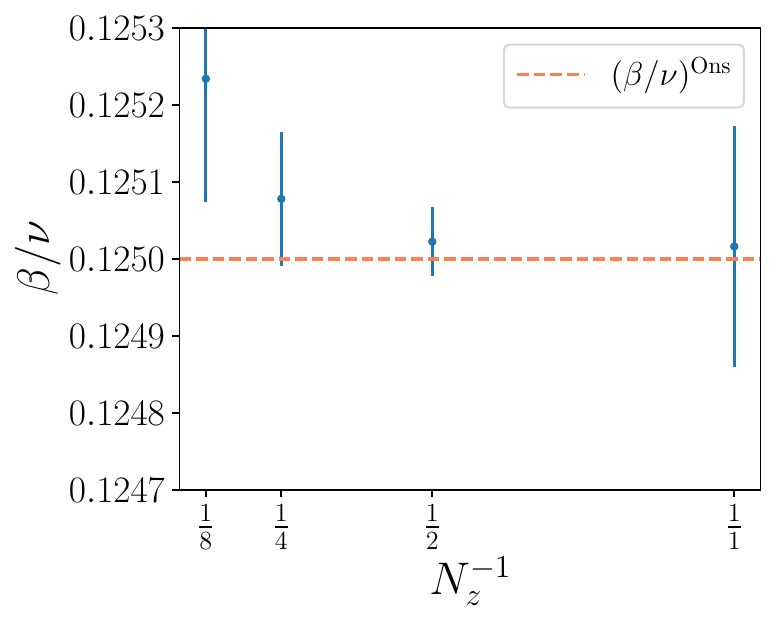}
     \hspace*{1mm}
     \includegraphics[width=0.45\linewidth]{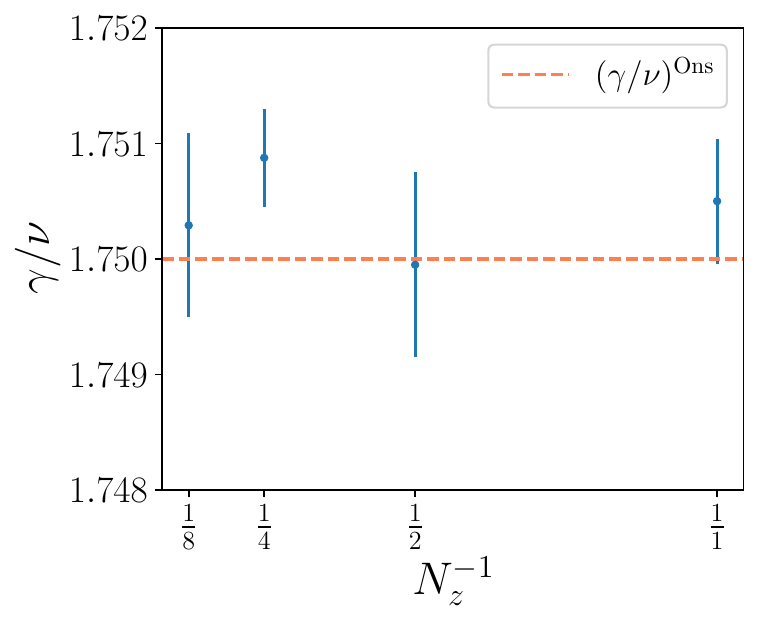}
     \hspace*{2mm}
     \includegraphics[width=0.45\linewidth]{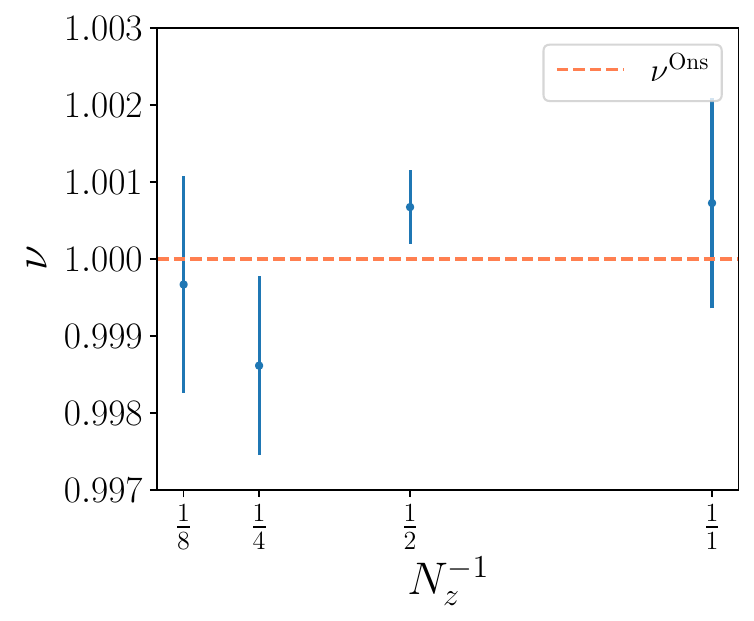}
     \includegraphics[width=0.46\linewidth]{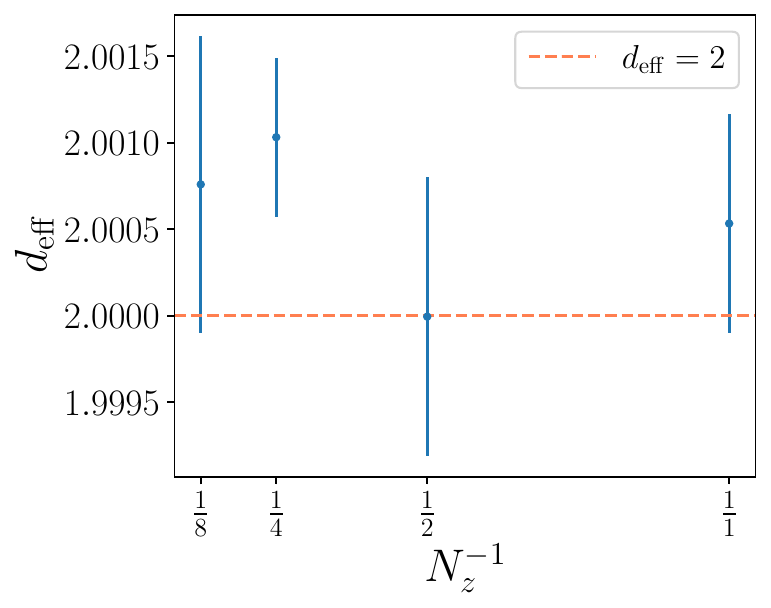}
     \vspace*{-2mm}
  \caption{Critical exponents $\beta/\nu$, $\gamma/\nu$, $\nu$ and $d_{\mathrm{eff}}$ (top left to bottom right) as a function of $N_z^{-1}$. The analytical values of the 2D Ising model are shown as dashed lines.}\label{fig:final_results}
\end{figure}

To get an estimator for the critical exponents $\beta/\nu$, $\gamma/\nu$ and $\nu$, we fit the finite size scaling relations (eq.~\ref{eq:scaling-relations}) to our numerical data. Fig.~\ref{fig:exp_scan} shows estimates for the exponents $\gamma/\nu$ and $\nu$ for $N_z=4$ and $N_z=8$, as a function of the minimal spatial lattice size included in the fit $(L_{\mathrm{min}})$. The estimators seem to rapidly decrease until $L_{\mathrm{min}}=448$ and $L_\mathrm{min}=256$, respectively, where they reach a plateau (with our error bars). Repeating the analysis in the same fashion for all other exponents and values of $N_z$, and calculating the effective dimension
\begin{equation}\label{eq:d_eff}
    d_{\mathrm{eff}}=\frac{2\beta+\gamma}{\nu},
\end{equation}
we collect our results in Tab.~\ref{tab:final_results} and display them (as a function of $1/N_z$) in Fig.~\ref{fig:final_results}.
We find no dependence of the critical exponents on $N_z$; in fact our results for $\beta/\nu$, $\gamma/\nu$ and $\nu$ at any given $N_z$ are consistent with the analytically known scaling exponents of the 2D Ising model.

\begin{table}[hbt!]
    \centering
    \begin{tabular}{|c|c|c|c|c|c|}
    \hline
    $N_z$ & $\Tilde{J_c}$ & $\beta/\nu$ & $\gamma/\nu$ & $\nu$ & $d_{\text{eff}}$ \\
    \hline
    1 & $0.440\,686\,94(44)$  & $0.125\,02(16)$  & $1.750\,50(54)$ & $1.000\,3(15)$  & $2.000\,53(63)$ \\
    2 & $0.276\,032\,19(13)$  & $0.125\,023(45)$ & $1.749\,95(80)$ & $1.000\,39(59)$ & $1.999\,99(81)$\\
    4 & $0.236\,027\,75(14)$  & $0.125\,078(87)$ & $1.750\,88(42)$ & $0.998\,8(15)$  & $2.001\,03(46)$\\
    8 & $0.226\,103\,634(93)$ & $0.125\,23(16)$  & $1.750\,29(80)$ & $0.999\,7(12)$  & $2.000\,76(86)$ \\
    \hline
    3D & $0.221\,654\,94(49)$ & $0.519\,3(13)$ & $1.963\,2(50)$ & $0.628\,75(82)$ & $3.001\,9(61)$\\
    \hline
    \end{tabular}
    \caption{Critical exponents $\beta/\nu$, $\gamma/\nu$ and the effective dimension
    $d_\mathrm{eff}$ for various choices of $N_z$. The analytic values for the 2D Ising model are $\tilde{J}_c=0.440\,686\,79\cdots$, $\beta/\nu=0.125$, $\gamma/\nu=1.75$, $\nu=1$ whereupon $d_{\mathrm{eff}}=2$.}
    \label{tab:final_results}
\end{table}

\newpage
\section{Conclusions}
We have studied a 3D Ising model with a mixture of the Metropolis algorithm and the Wolff cluster flipping algorithm. Data analysis has been performed by means of histogram reweighting and finite size scaling techniques. We considered the case where $N_{x,y,z}\to\infty$ as well as the dimensionally reduced case where $L=N_x=N_y\to\infty$ with fixed $N_z=1,2,4,8$. Using a wide range of system sizes, we have obtained results for $\tilde{J}_c$, $\beta/\nu$, $\gamma/\nu$, $\nu$ and $d_{\mathrm{eff}}$. Our 3D results are compatible with the latest results of A.~M.~Ferrenberg, J.~Xu and D.~P.~Landau \cite{Ferrenberg:2018zst}. Regarding $\tilde{J}_c$ and $\nu$ our dimensionally reduced results are compatible with (though more precise than) the results of M.~Caselle and M.~Hasenbusch \cite{Caselle:1995wn}. Regarding $\beta/\nu$ and $\gamma/\nu$ we are unaware of a publication with similarly accurate results at fixed $N_z$ to check against. In any case our results suggest that all $\tilde{J}(N_z)$ lie on a smooth curve which connects the analytically known value in 2D ($N_z=1$) with the well known value in 3D ($N_z\to\infty$). For any finite $N_z$ our critical exponents suggest that the model is still in the 2D Ising model universality class.

\end{document}